\documentclass[]{spie}  %>>> use for US letter paper
\usepackage{amsmath,amsfonts,amssymb}
\usepackage{graphicx}
\usepackage[colorlinks=true, allcolors=blue]{hyperref}
\usepackage[super,sort,compress]{natbib} %added by lucas to prevent dot before citation
\usepackage{textcomp}
\usepackage{float} % added by Willeke

\title{Spectropolarimetry of life: \\airborne measurements from a hot air balloon}

\author[a,b]{Willeke Mulder}
\author[c]{C. H. Lucas Patty}
\author[c]{Stefano Spadaccia}
\author[c]{Antoine Pommerol}
\author[d]{Brice-Olivier Demory}
\author[a,e]{Christoph U. Keller}
\author[d,f]{Jonas G. Kühn}
\author[a]{Frans Snik}
\author[b]{Daphne M. Stam}
\affil[a]{Leiden Observatory, Leiden University, PO Box 9513, 2300 RA Leiden, the Netherlands;}
\affil[b]{Faculty of Aerospace Engineering, Delft University of Technology, Kluyverweg 1, 2629 HS Delft, The Netherlands;}
\affil[c]{Physikalisches Institut, Universit\"at Bern, Sidlerstrasse 5, 3012 Bern, Switzerland;}
\affil[d] {Center for Space and Habitability, Universit\"at Bern, Gesellschaftsstrasse 6,  3012 Bern, Switzerland}
\affil[e]{Lowell Observatory, 1400 W Mars Hill Rd, Flagstaff, AZ 86001, USA;}
\affil[f] {D\'epartement d'Astronomie, Universit\'e de Gen\`eve, 1290 Versoix, Switzerland}

\authorinfo{Further author information: (Send correspondence to W.M.)\\W.M.: E-mail: wmulder@strw.leidenuniv.nl, www.willekemulder.com}

\begin{document} 
\maketitle

%%%%%%%%%%%%%%%%%%%%%%%%%%%%%%%%%%%%%%%%%%%%%%%%%%%%%%%%%%%%%%%%%%%%%%%%%%
\begin{abstract} % not more than 250 words

Does life exist outside our Solar System? A first step towards searching for life outside our Solar System is detecting life on Earth by using remote sensing applications. 
One powerful and unambiguous biosignature is the circular polarization resulting from the homochirality of biotic molecules and systems. We aim to investigate the possibility of identifying and characterizing life on Earth by using airborne spectropolarimetric observations from a hot air balloon during our field campaign in Switzerland, May 2022. 

In this work we present the optical-setup and the data obtained from aerial circular spectropolarimetric measurements of farmland, forests, lakes and urban sites. We make use of the well-calibrated FlyPol instrument that measures the fractionally induced circular polarization ($V/I$) of (reflected) light with a sensitivity of $<10^{-4}$. The instrument operates in the visible spectrum, ranging from 400 to 900 ~nm. We demonstrate the possibility to distinguish biotic from abiotic features using circular polarization spectra and additional broadband linear polarization information. We review the performance of our optical-setup and discuss potential improvements. This sets the requirements on how to perform future airborne spectropolarimetric measurements of the Earth's surface features from several elevations.
\end{abstract}


%%%%%%%%%%%%%%%%%%%%%%%%%%%%%%%%%%%%%%%%%%%%%%%%%%%%%%%%%%%%%%%%%%%%%%%%%%
\section{INTRODUCTION}
\label{sec:intro}  

The remote-sensing of the Earth gives us indispensable information for our preparations to search for extraterrestrial life. Well-known examples of biosignatures that can be identified in the reflected sunlight are atmospheric constituents such as O$_2$ and spectral signatures of vegetation, such as the green bump and the red edge (see Seager et al. (2005)\cite{Seager05} for a detailed description of the red edge). In addition, spectropolarimetry proves itself to be a robust remote-sensing tool. Complementary to the reflectance, it carries additional and unique information, such as surface roughness or particle size of the scatterers which helps us to look for and characterize bio-signatures. Spectropolarimetry also provides unique biosignatures itself, such as the linear polarization resulting from the O2-A band \cite{Stam99,Fauchez17} and from vegetation\cite{Vanderbilt85,Vanderbilt91}, as well as circular polarization resulting from the homochirality of biotic systems\cite{Patty19,Gimenez19}. % linear polarisation is already used at the moment to indentify the health status of plants

Homochirality is what we refer to as the single handedness of chiral molecules. Terrestrial chiral molecules that are utilized by life, e.g. amino acids, sugars and the biological polymers they construct, almost always exists in either left-handed or right-handed forms, and are therefore homochiral. This is in contrast to abiotic chemistry, which produces equal amounts of these mirror forms, resulting in a racemic mixture. Circular polarization originates from the differential absorption by homochiral molecules and macromolecular structures. As biochemical homochirality is essential for life and thought to be a universal property of life, the circular polarization life produces constitutes an unambiguous bio-signature \cite{Patty18}. Therefore, circular polarization promises to be a powerful tool for the remote-sensing of biotic matter on Earth \cite{Wolstencroft04,Patty17,Patty19,Patty21} and beyond \cite{Sparks05,Patty18}. 

In general, direct light emitted by a solar type star is virtually unpolarized when integrated across its stellar disk \cite{Kemp87}. Since linear polarization is produced by interaction of light with surfaces and particles, whenever Sunlight interacts with the Earth's atmosphere or surface, it usually becomes (partly) linearly polarized. As such, we do find an abundance of polarized light by looking at the Earth's atmosphere or surface. Various of these polarizing scattering mechanisms, like Rayleigh scattering and reflections at air-water interfaces, are well understood\cite{Cronin11}. Despite all this knowledge, the interpretation of remotely sensed linear polarization data of the real world is challenging as it involves e.g. depolarization effects, varying atmospheric aerosol concentrations and a diversity of (cloud)particle faces, shapes and orientations. 

Circularly polarized light is much more scarce in nature. %In the universe, circular polarization can originate from the single scattering of light on magnetically aligned non-spherical dust particles in Star-Formation Regions \cite{Bailey98}. 
It can be produced through multiple scattering processes. Single scattering processed usually generate linearly polarized light, after which a second scattering event with atmospheric aerosols can produce circularly polarized light. We refer to Gasso et al. (2022)\cite{Gasso22} for an elaborate summary of circular polarization due to atmospheric aerosols. In addition to multiple scattering processes, circularly polarized light can be produced by the homochirality of biotic systems\cite{Patty19,Patty21}. 

The development of theoretical scattering models including both linear and circular polarization is essential to understand all information coming from remotely sensed spectropolarimetric data. There exist various extensive spectropolarimetric models of Earth-like (exo)planets featuring realistic atmosphere profiles\cite{stam08}\, realistic cloud parameters\cite{Groot20}, wind-ruffled oceans with sea foam and shadows of the waves\cite{Trees22} and multiple surface reflection scattering matrices based on bidirectional reflectance functions and characteristic (wavelength-dependent) surface albedos\cite{stam08}. For the latter, surface albedos for many different natural and man-made materials are provided by libraries such as the ASTER\cite{Baldridge09} and the ECOSTRESS spectral library\cite{Meerdink19}.

\begin{figure} [ht]
   \begin{center}
   \begin{tabular}{c} 
   \includegraphics[width=0.45\textwidth]{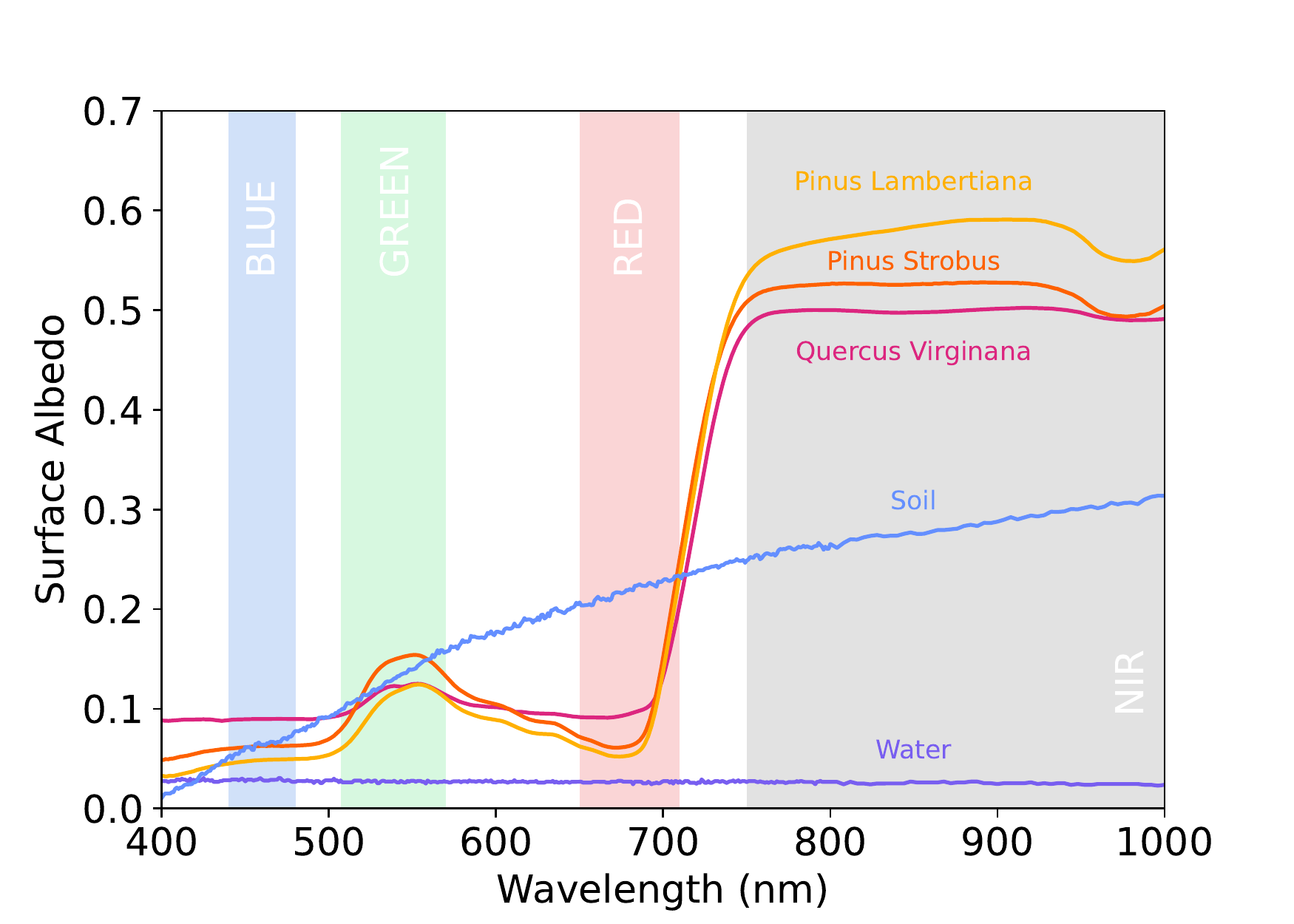}
	\end{tabular}
	\end{center}
   \caption{The wavelength dependent surface albedos of 3 different types of trees, (i) Quercus Virginana, (ii) Pinus Strobus, and (iii) Pinus Lambertiana, and soil and water (These five albedos have been taken from the ECOSTRESS Spectral Library\cite{Meerdink19}). \label{fig: albedo}  }
\end{figure}

Surface albedos can be used for surface identification. For example, the spectral shapes of vegetation features can be easily distinguished from bare soils\cite{Liang02}, see Figure \ref{fig: albedo}. Albedos might vary over the year due to the change of seasons, change in vegetation characteristics or their moistness. The vegetation albedo spectra share the following characteristics: (i) absorption bands of chlorophyll around 435-485 nm and around 645-685 nm, (ii) a high albedo at wavelengths longer than 700 nm, which is also referred to as the red edge and (iii) the absorption of light due to intracellular liquid water causing slight dips around 0.97, 1.15, 1.45, and 1.92 µm \cite{Hedge15}. The definition of surface albedo is commonly used in the field of astronomy and climate research. For simplicity, we will refer to the surface albedo as the (surface) reflectance. 

Even though there has been an extensive advancement of the models over the years, there is still room for improvement. The surface models derived from scattering matrices based on the surface reflectance serve their purpose very well when considering a planetary disk-integrated signal. Unfortunately, they do not include circular polarization signals induced by vegetation features. In general, and especially in nature, circular polarization signals are very faint compared to those of linear polarization: the difference can easily be three orders of magnitude. However, being a powerful tool for the remote-sensing of biotic matter, we do want to investigate the possibility of adding circular polarization to existing surface models. 

Remote-sensing of (circularly) polarized biosignatures has received a lot of interest in recent years \cite{Patty19,Patty21,Snik19,Sparks20}. Data acquired through remote-sensing offers crucial information for preparation for the design and development of the next generation of space-based spectropolarimeters. We follow up upon the results of Patty et al. (2021)\cite{Patty21}. They used a helicopter to perform airborne measurements. In this work, we are using a stable and cheaper airborne platform: a hot air balloon. One of the biggest advantages of a hot air balloon is that our proximity to our instrument during the flight allows us make on the fly adjustments to the optical-setup.  
We measure the light that is reflected by the surface from the moving balloon basket, hence, we are limited to a single illumination angle, viewing angle and elevation per surface scene. Therefore, our main focus lies with measuring the circular polarization spectra from landscapes and their use in surface identification. In addition to circular polarization, we continuously record four individual linear polarization states using an ordinary polarization camera.  %This technology was first patented in 1995\cite{Wolff95}, that lead to more practical implementations and technology advances since 2009\cite{Zhao09,Gruev10,Myhre12} in medical applications\cite{Momeni06,Tokuda09}, interferometry and remote-sensing applications\cite{Tyo96,Tyo06}. Studies use aluminum nanowire optical filters\cite{Gruev10}, combinations of dichroic dye and liquid crystal polymers\cite{Myhre12},  This proofs to be a useful tool for remote-sensing applications\cite{Tyo06}. 
We investigate the contribution and practicality of using a polarization camera for future airborne field campaigns. 

This paper has the following structure. In Section \ref{sec: methods}, we describe the basics of polarimetry and the concept of the Normalized Difference Vegetation Index (NDVI) to distinguish between different surface types. In Section \ref{sec: instrument}, we describe our instrumental set-up that was designed to measure the fractional circular polarized reflection of the Earth from a balloon. In Section \ref{sec: results}, we summarize the main results from the airborne measurements. At the end, we finalize this paper with our conclusions and discussion in Section \ref{sec: conclusion}.

%-------------------------------------------------------
%-------------------------------------------------------

%%%%%%%%%%%%%%%%%%%%%%%%%%%%%%%%%%%%%%%%%%%%%%%%%%%%%%%%%%%%%%%%%%%%%%%%%%
\section{Methods}
\label{sec: methods}

%-------------------------------------------------------------------------
\subsection{Polarization}

Polarization is generally described with a Stokes vector $\textbf{S} = (I,Q,U,V)$, where $I$ is the total intensity, $Q$ and $U$ are the 
linearly polarized intensities and $V$ the circularly polarized intensity. The complete polarization states are described in terms of the normalized Stokes parameters $Q/I$, $U/I$ and $V/I$ (each ranging from -1 to 1), where $Q/I$ denotes the difference between linearly polarized 
intensities normal and parallel to the plane of scattering, $U/I$ denotes the difference between +45$^\circ$ and -45$^\circ$ to the plane of scattering, and $V/I$ denotes the difference between right-handed and left-handed circularly polarized light.

The linear Stokes parameters $Q$ and $U$ can be combined into the (dimensionless) Degree of Linear Polarization, DoLP. The DoLP of a surface
indicates the fraction of the reflected light that is linearly polarized. It can provide us with essential information about land surface characteristics.  The Angle of Linear Polarization, AoLP, contains additional information  related to surface or material properties. 
The DoLP and AoLP are expressed as:
\begin{equation} \label{eqn: dolp&aolp}
    \text{DoLP} = \frac{\sqrt{Q^2+U^2}}{I}\vspace{5pt}; \hspace{10pt}
    \text{AoLP} = \frac{1}{2} \tan^{-1}\left(\frac{U}{Q}\right).
\end{equation}

%-------------------------------------------------------------------------
\subsection{The NDVI}

Using satellite imaging techniques, we can measure the reflected light spectra of different types of Earth surfaces. However, calculating an accurate widespread surface reflectance can be complex. Instead, we choose to calculate the Normalized Difference Vegetation Index (NDVI)\cite{Rouse74} for individual measurements to distinguish between different surface types.
We calculate the NDVI as according to Patty et al. (2021)\cite{Patty21}
\begin{equation}
    \text{NDVI} = \frac{I_{\text{NIR}}-I_\text{{R}}}{I_{\text{NIR}}+I_{\text{R}}},
\end{equation}
where $I_{\text{R}}$ is the mean of the reflectance from 
$650 \mathrm{nm}$ to $710 \mathrm{nm}$ and $I_{\text{NIR}}$ is the  mean from $750 \mathrm{nm}$ to $780 \mathrm{nm}$.

This NDVI ranges from -1.0 to +1.0, where the negative values are likely to identify water.  NDVI values close to +1 have a large possibility to indicate green vegetation, as this absorbs solar radiation for photosynthesis in the spectral region from 400 to 700~nm. It reflects radiation in the NIR-region of the solar spectrum (the Red Edge). For green vegetation, $I_{\text{NIR}}$ is thus much larger than $I_{\text{R}}$ and the NDVI is close to 1.0.  An NDVI close to zero indicates an absence of green vegetation, for example, an urban area with roofs, roads, and/or concrete. 

%%%%%%%%%%%%%%%%%%%%%%%%%%%%%%%%%%%%%%%%%%%%%%%%%%%%%%%%%%%%%%%%%%%%%%%%%%
\section{Instrumental set-up}
\label{sec: instrument}

Circular spectropolarimetric measurements were performed with the FlyPol\cite{Patty21} instrument. FlyPol is a spectropolarimeter, based on the TreePol design\cite{Patty17,Patty19}, that uses fast temporal polarization modulation to obtain the fractionally induced circular polarization ($V/I$) of the observed light as a function of wavelength from 400 to 900~nm. The instrument sensitivity ($< 10^4$) and accuracy ($< 10^3$) is high enough to measure the circular polarization induced by vegetation. FlyPol improves its stability in the field by actively controlling the temperature of its optics and electronics. Its angular field of view is approximately 1.2°\cite{Patty21}.

All airborne measurements presented in this paper were obtained during one flight in an Ultramagic hot air balloon carrying a so-called T-partition basket with a capacity of 11 passengers including the pilot. 
The volume of this hot air balloon is relatively large, 8500 cubic meters, which allowed us to take FlyPol and four scientists to perform the measurements. The size of the entire basket was $1.2\times1.75\times3.05$ $(\text{h}\times\text{w}\times\text{l})$ meters. A single partition was large enough ($\sim 0.80\times2.00$ meters) to fit the tripod
to which FlyPol was connected. All electronics could be stored beneath the tripod, leaving enough space in the partition for two persons. 

\begin{figure} [ht!]
   \begin{center}
   \begin{tabular}{c} 
   \includegraphics[width=0.95\textwidth]{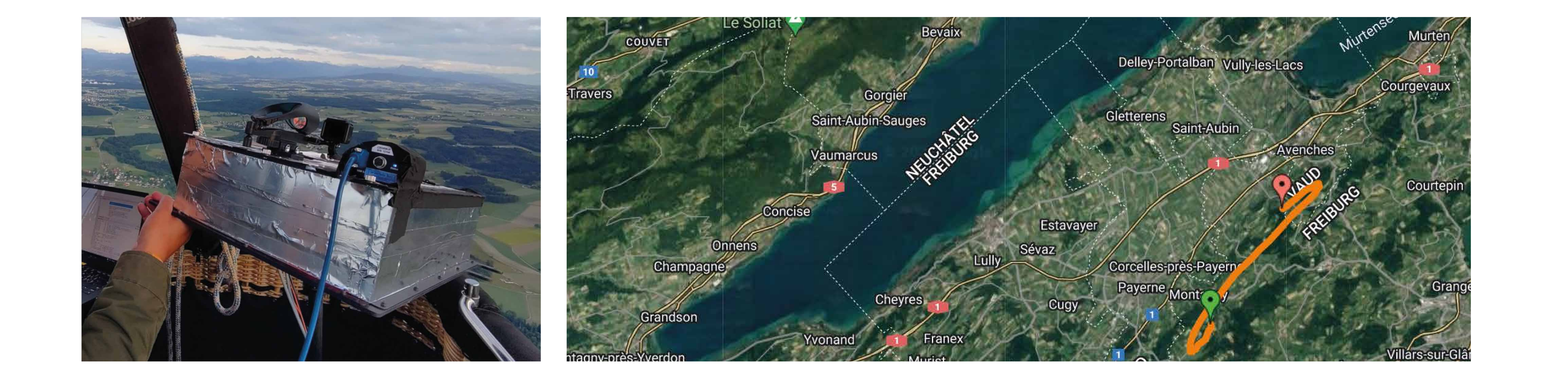}
	\end{tabular}
	\end{center}
   \caption[example] 
   { \label{fig: pointing-cameras} [left] Photo of the FlyPol instrument in the basket of the hot air balloon. A parallax-free telescope pointer, a GoPro and polarization camera are connected to the top of the instrument. They collect and store additional pointing information. [right] Map of the flight trajectory on Monday, May 30, 2022. Satellite image from Google Maps.}
\end{figure}

FlyPol was oriented using a calibrated parallax-free telescope pointer\cite{Patty21}. A GoPro HERO7 reference camera and a broadband Thorlabs' Polarization-Sensitive Kiralux$^{\tiny{\text{\textregistered}}}$ camera were aligned with the pointer and mounted on top of FlyPol's instrument casing, see Figure \ref{fig: pointing-cameras}. The polarization camera features a 5.0 MP monochrome CMOS sensor with a wire grid polarizer array. The array consists of a repeating pattern of four linear polarizers with transmission orientation axes of 0°, 45°, 90° and 135°, respectively.  
A 'super-pixel' calibration algorithm\cite{Gimenez19,Lane22} corrects the images for dark noise, flat fielding, and the optical imperfections of the Polarization Filter Array. In this way, we acquired pointing and additional information regarding the observed areas, which is insightful when analyzing the data. 

%-------------------------------------------------------
%-------------------------------------------------------

Two individual GPS trackers were used to record the flight trajectory, see Figure \ref{fig: pointing-cameras}. The balloon flew over the Broye district in the canton of Fribourg, Switzerland. This region lies on an elevated, relatively flat plain that is called the Swiss Plateau. The canton is predominantly rural, featuring farm sites, valleys and small forests. With an altitude of 2389~m, the Vanil Noir is the highest mountain in the canton. This is about 200~m lower than the highest altitude that we reached during our flight. From hereon, we refer to the elevation of the balloon, which is the difference between measured GPS altitude in the air and the local ground-level altitude. The highest elevation was just below 2000 m. The wind determines both the flight speed and direction. At altitudes $< 1500$~m we were heading South with a ground speed of $\sim$ 10 km/h, while the winds at altitudes $> 1500$~m droves us North with a ground speed of $\sim$27 km/h. The pilot could turn the orientation of the balloon, and thus the basket, in the timeframe between the end of the take-off and the start of the landing. 

\begin{figure} [ht]
   \begin{center}
   \begin{tabular}{c} 
   \includegraphics[width=0.95\textwidth]{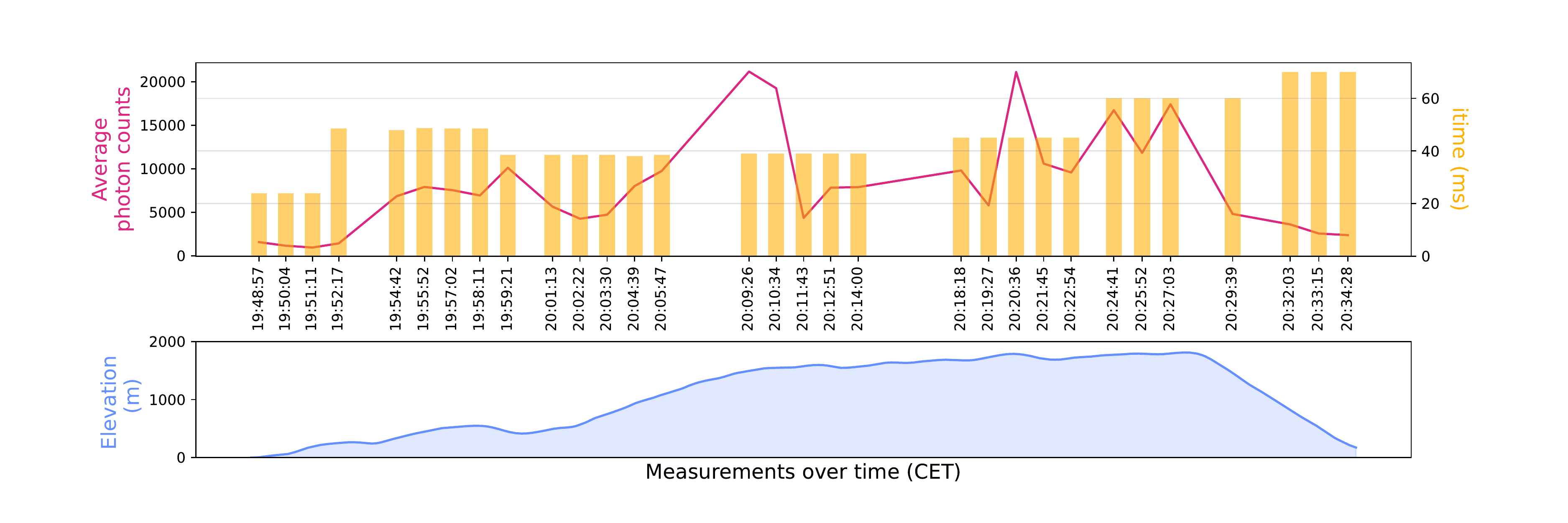}
	\end{tabular}
	\end{center}
   \caption{ [Top] The average raw photon count for FlyPol's spectrometer (in the wavelength range of 400-900 nm) and the integration time for each individual scene. [Bottom] The elevation of the balloon 
   (thus its altitude above the ground) as measured during the flight.}
   { \label{fig: itimevscounts} 
}
   \end{figure}

In total, 31 measurements were obtained under clear sky conditions on May 30, 2022, from 19:45 to 20:45 CEST, just before sunset (21:20 CEST). We choose to fly during the evening hours, as hot air balloons are only able to fly within the time-span of 3 hours before sunset until 3 hours after sunrise. In this work, we refer to a single, continuous measurement as a `scene'. During the flight, the solar zenith and azimuth angle were approximately 86°-90° and 300°-310°, respectively. Throughout the flight, FlyPol's integration times were varied per scene to obtain the highest average photon count possible while preventing saturation, see Figure \ref{fig: itimevscounts}. For the first and last measurements, the average photon count ($< 10000$ counts) is low compared to the later measurements. While approaching sunset, the integration time increased as the photon counts strongly decreased with time. At the same time, we adjusted the integration times of the polarization camera, as the auto-exposure mode caused saturation of the first set of images. The polarization camera was set to record one exposure every 5 seconds. Unfortunately, adjusting the integration times took at least a minute, leaving holes in the dataset. The reference camera took an individual automatic shutter image every 30 seconds\footnote{30 seconds is the bare minimum for automatic shutter imaging with a GoPro HERO7.}.

%-------------------------------------------------------
%-------------------------------------------------------

%%%%%%%%%%%%%%%%%%%%%%%%%%%%%%%%%%%%%%%%%%%%%%%%%%%%%%%%%%%%%%%%%%%%%%%%%%
\section{Results}
\label{sec: results}

\begin{figure} [ht]
   \begin{center}
   \begin{tabular}{c} 
   \includegraphics[width=0.87\textwidth]{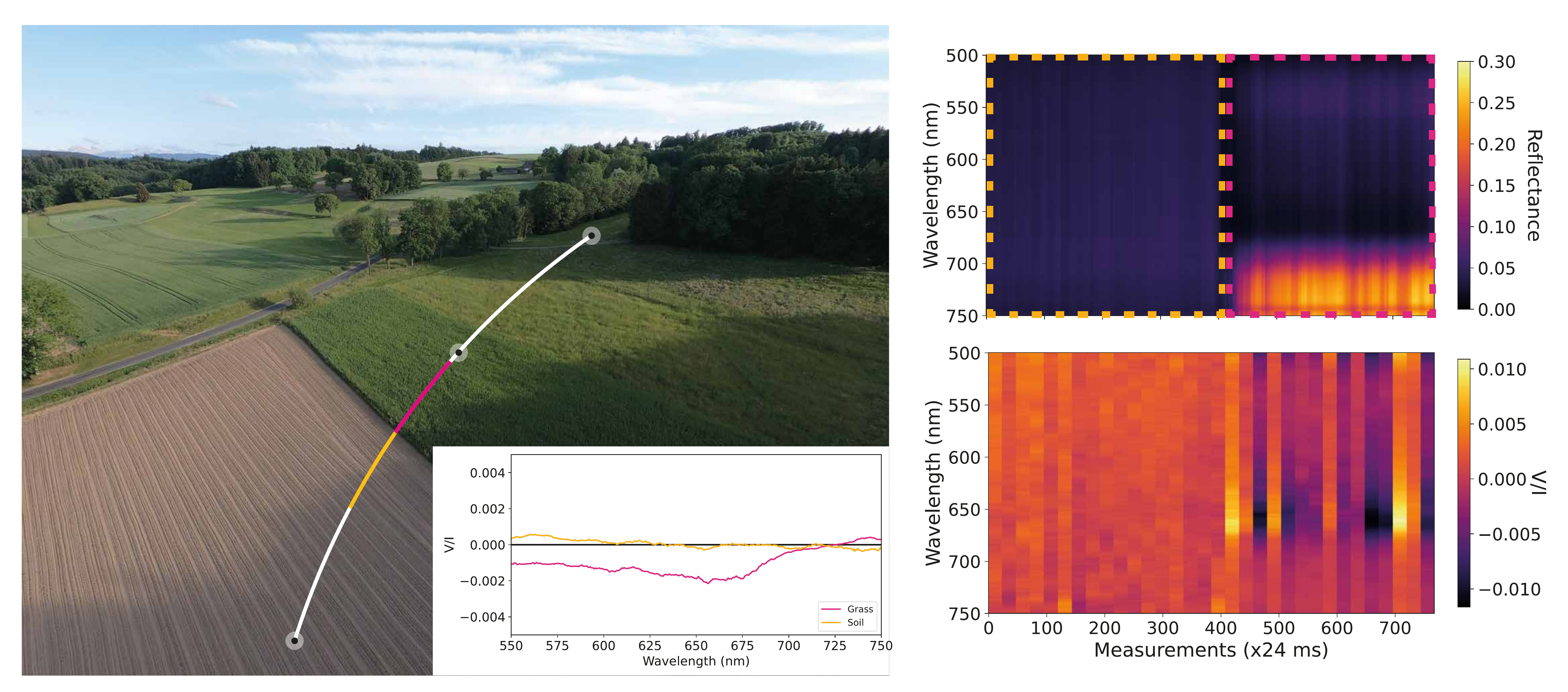}
	\end{tabular}
	\end{center}
       \caption{[Left] 
       Single measurement scene during take-off, at 19:48:57 CEST at $\sim$~20~m elevation and $\sim$~5~km/h ground speed, near Montagny, while flying in the SW direction. The transparent black/white dots mark the pointing of FlyPol for three sequential measurements with a 30 seconds time interval. The inset indicates the circular polarization spectrum averaged over the individual soil and grass scenes in yellow and pink, respectively. The transparent areas indicate the standard deviation of the averaged measurements. 
       [Top right] Qualitative reflectance data for the trajectory per wavelength, over time. [Bottom right] Circular polarization
       $V/I$ for the trajectory per wavelength, over time.}
\label{fig: observation_194857_G0013896}
\end{figure}

In Figure \ref{fig: observation_194857_G0013896}, we show the results of a scene observed during take-off, at an elevation of about 20~m. 
The reference photo on the left contains three transparent black/white dots that mark the central pointing of FlyPol for three sequential measurements with a 30 seconds time interval. The line connecting the dots indicates the likely intermediate pointing trajectory. The panels on the right show the qualitative reflectance and circular polarization $V/I$ for the trajectory per wavelength, over time. 
Both plots show a very clear separation between the reflected light signals of the soil and the grass after $\sim400$ measurements. The soil shows a spectrally flat reflectance and a negligible circular polarization. The grass shows a strong increase in the reflectance in the Red Edge, around 710~nm. The inset in the left image reveals the circular polarization spectrum averaged over time for the soil and grass features on the right. The grass spectrum reveals a negative band with a minimum at 675 nm with a magnitude of $V/I_{\text{min}} = 2.0\times 10^{-3}$.

%-------------------------------------------------------
%-------------------------------------------------------
%-------------------------------------------------------
\begin{figure} [ht]
   \begin{center}
   \begin{tabular}{c}  \vspace{-10pt}
   \includegraphics[width=0.87\textwidth]{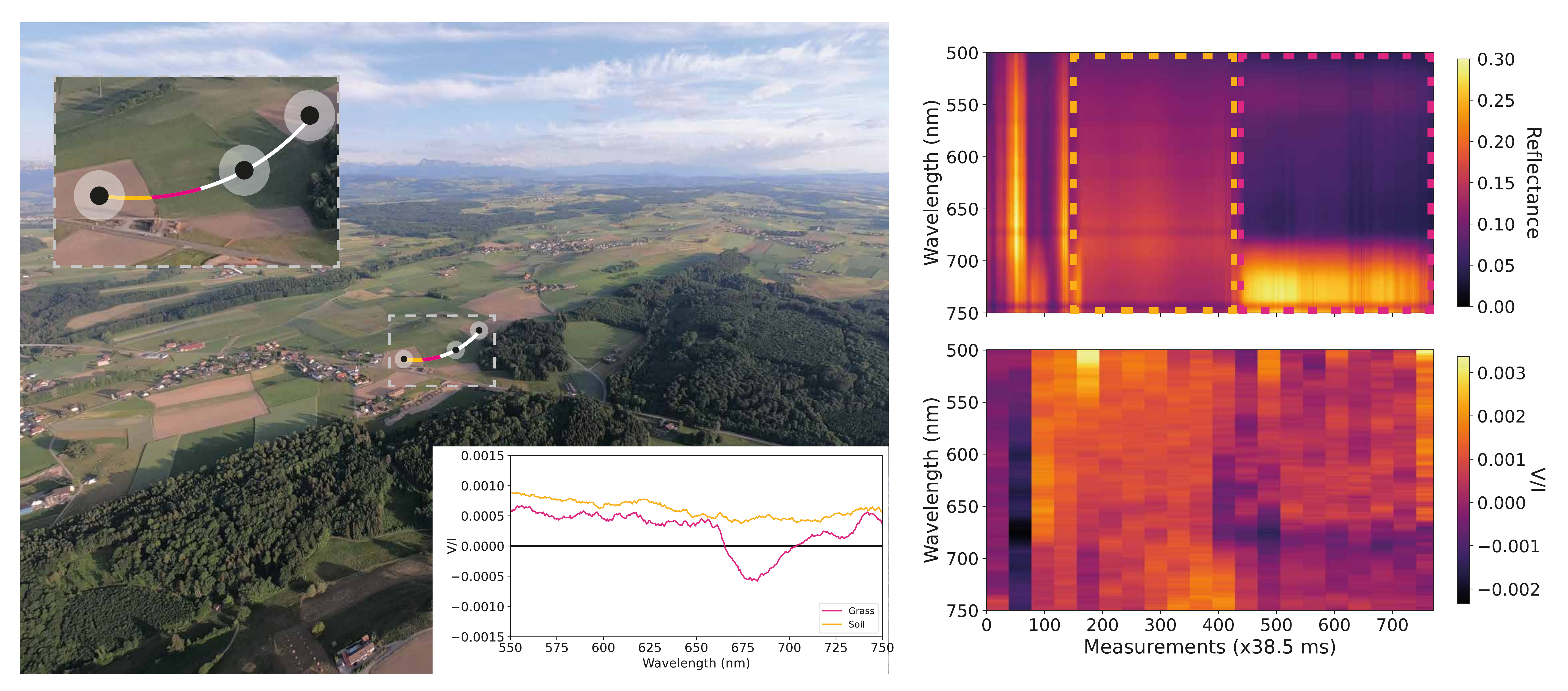}
	\end{tabular}
	\end{center}
    \caption{Similar to Figure~\ref{fig: observation_194857_G0013896},
    except at 20:01:13 CEST, with $\sim$~650 m elevation and $\sim$8.5 km/h ground speed.}
    \label{fig: observation_200113_G0013921} 
   \end{figure} 
%-------------------------------------------------------

In Figure \ref{fig: observation_200113_G0013921} we show the results for a scene, pointing at a rural area from an elevation of $\sim$~650~m. The yellow and pink lines in the white line trajectory match the location of the in yellow and pink highlighted qualitative reflectance data for the trajectory per wavelength, over time. We were able to identify the two different surface types using the reflectance and the NDVI. 
Just as for the reflectance in Figure \ref{fig: observation_194857_G0013896}, there is an increase in the reflectance around 675~nm due to the Red Edge. The measured $V/I$ is smaller than for the previous scene.

\begin{figure} [H]
   \begin{center}
   \begin{tabular}{c} 
   \includegraphics[width=0.65\textwidth]{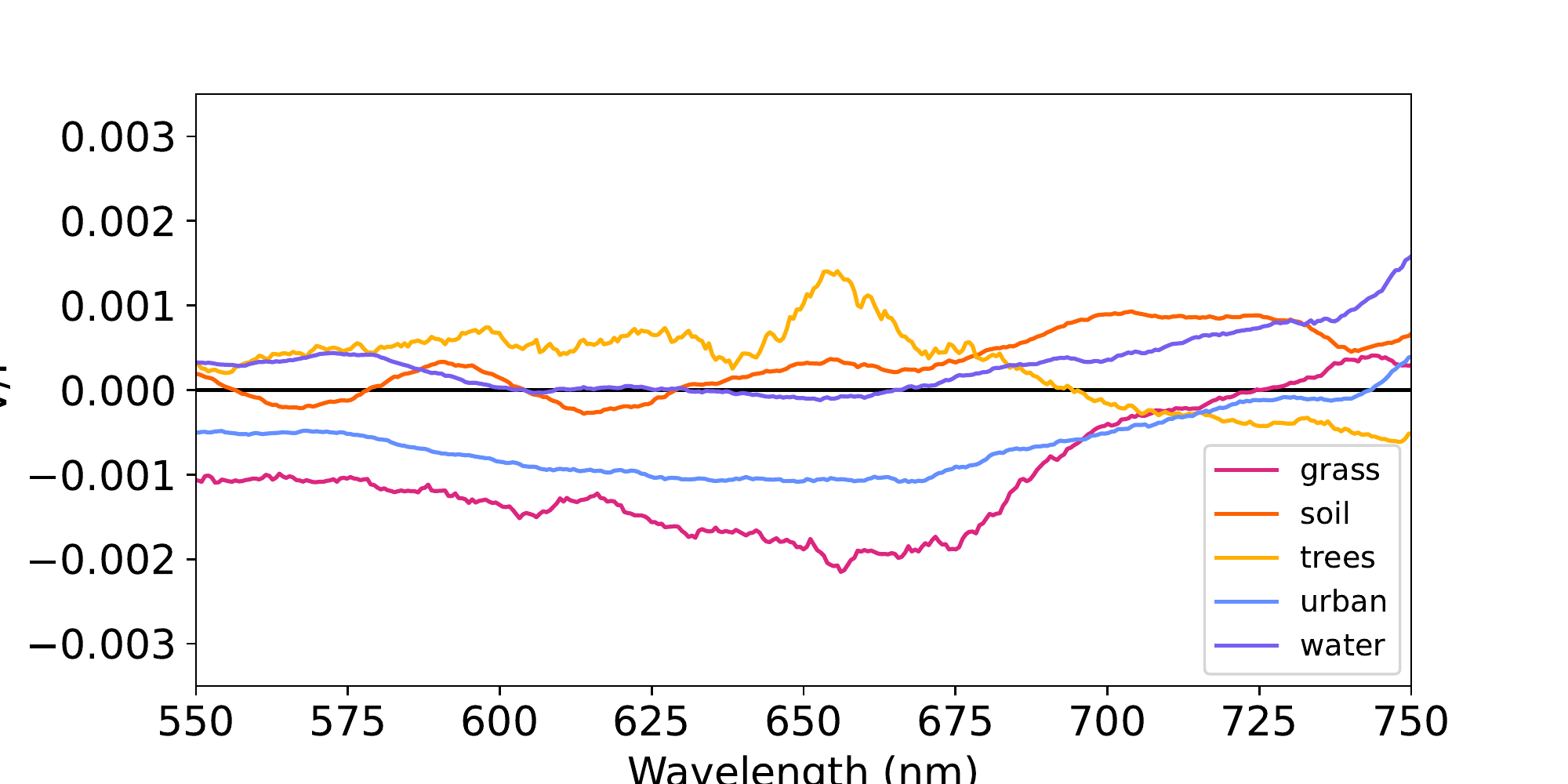}
	\end{tabular}
	\end{center}
   \caption{Circular polarization ($V/I$) spectra for different surface scenes: grass, soil, trees, urban, and (lake) water.}
   \label{fig: landscapesVI} 
   \end{figure}

Using the NDVI values and the reference photos for the individual scenes, we were able to differentiate between five surface types; grass, soil, trees, urban, and water. The spectra presented in Figure \ref{fig: landscapesVI} are time averaged $V/I$ spectra. The scenes that feature vegetation (grass, trees) are easily distinguishable from the others, due to their high ($> 0.75$) NDVI values. As explained earlier, this is due to the relative strong reflection of near-infrared light and absorption of red light by the chlorophyll molecules. The circular polarization spectra of trees had a positive polarization band of $V/I=1.0\times^{-3}$ around 650 nm. The grass has a negative band with a minimum of $V/I=2.0\times^{-3}$ around 660 nm. Beyond $\sim$~675~nm, $V/I$ decreases. As expected, scenes that feature the soil, urban and water do not show significant circular polarization signals.

%-------------------------------------------------------
%-------------------------------------------------------

\begin{figure} [H]
   \begin{center}
   \begin{tabular}{c} 
   \includegraphics[width=0.95\textwidth]{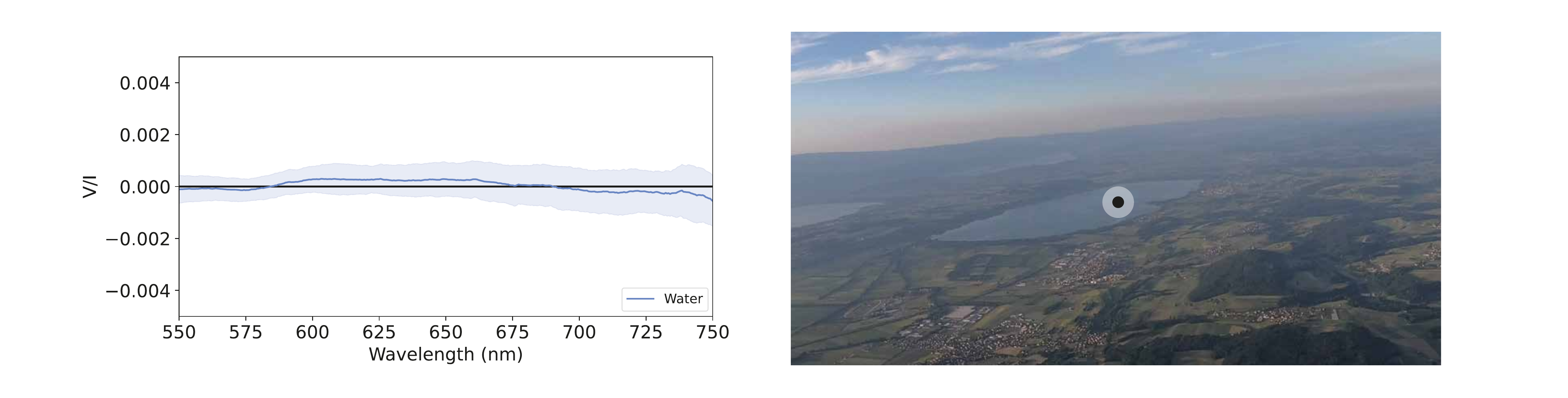}
	\end{tabular}
	\end{center}
   \caption{[Left] Time-averaged $V/I$ of water. [Right] Reference photo of the observed scene during take-off (20:12:51 CEST with $\sim$~1870~m elevation and $\sim$~25.0 km/h ground speed) flying towards Oleyres, in the NE direction while pointing towards Murtzensee. The black/white dot indicates the pointing of FlyPol.}
   \label{fig: observation_201251_G0013943}
\end{figure}

Patty et al. (2021)\cite{Patty21} flew over the lake Lac des Taillères and measured a circular polarization signal of $V/I = 1.1 \times 10^{-3}$, which indicates a possible presence of photosynthetic organisms like algae. We pointed FlyPol several times at the Murtzensee to try and detect a visible red edge as well. All our circular polarization spectra have a similar shape to the one presented in Figure~\ref{fig: observation_201251_G0013943}. We did not observe a circular polarization signal originating from surface water, which suggests an absence of biotic organisms. 

%-------------------------------------------------------
%-------------------------------------------------------

\begin{figure} [H]
   \begin{center}
   \begin{tabular}{c} 
   \includegraphics[width=0.95\textwidth]{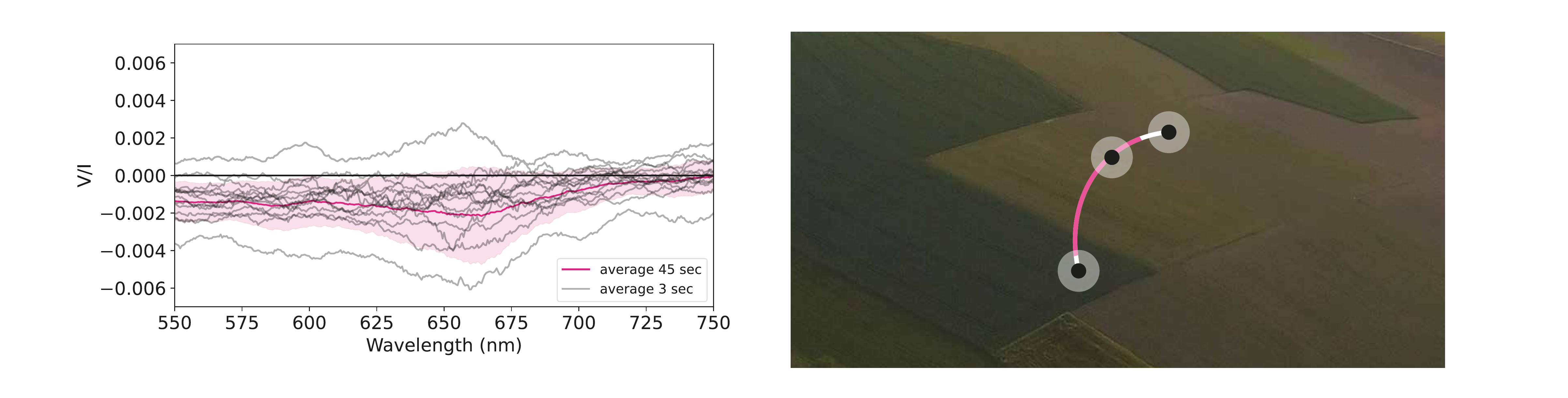}
	\end{tabular}
	\end{center}
   \caption{[Left] 15 subsequent three-second-averaged and the total 45-second-averaged circular polarization spectra. The scene captures farmland that features various types of grass and soil. [Right] Reference photo of a scene during take-off ((19:52:17 CEST at 
   $\sim$~790~m elevation and $\sim$~7.8 km/h ground speed) flying towards Ruisseau de Pra Laurent, in the S direction while pointing at farmland. The colors on the white trajectory indicate the measurement locations of the $V/I$ spectra displayed in the left figure.}
   \label{fig: observation_195217_healthyness_vegetation}
   \end{figure}

\begin{figure} [ht]
   \begin{center}
   \begin{tabular}{c} 
   \includegraphics[width=0.95\textwidth]{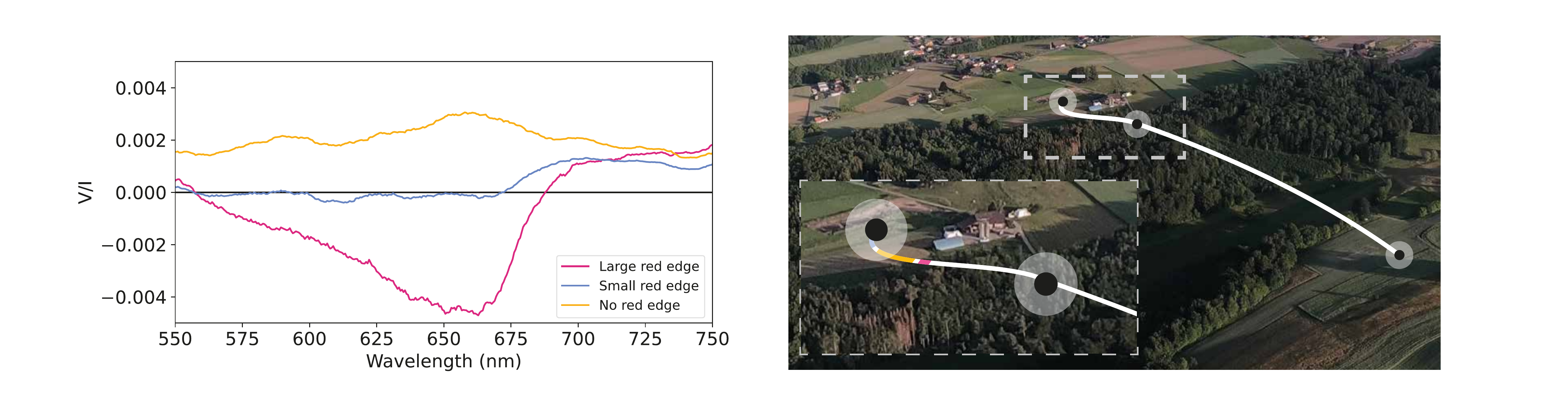}
	\end{tabular}
	\end{center}
   \caption{ [left] Time-averaged $V/I$ of a large, a small and no red edge effect, identified as trees, grass and soil respectively.  [right] Reference photo of a measurement scene during take-off (19:54:42 p.m. CEST with $\sim$540 m elevation and $\sim$~3.9 km/h ground speed) flying towards Mannens-Grandsivaz, flown in SW direction while pointing towards Marais Martin.} 
    \label{fig: observation_grass_trees_underdiscussion}  
   \end{figure}

The Figures~\ref{fig: observation_195217_healthyness_vegetation} and~\ref{fig: observation_grass_trees_underdiscussion} illustrate the difficulty of identifying multiple distinct surface features within one scene. Figure \ref{fig: observation_195217_healthyness_vegetation} contains 15~subsequent three-second duration circular polarization spectra originating from one observation scene while flying over farmland. The magnitude varies from $V/I_{\text{min}}= -9.0 \times^{-3}$ to $V/I_{\text{max}} = 2.5 \times^{-3}$ due to a large variety of observed grass and soil. This large variety makes it diffucult to designate an accurate source to all the individual lines. We end up with a signal of $V/I = 2.0 \times^{-3}$, see pink line in Figure \ref{fig: observation_195217_healthyness_vegetation}, when avereriging over all 15 spectra. The three spectra, large-, no- and small red edge, presented in Figure \ref{fig: observation_195217_healthyness_vegetation} are also originating from one observation scene. The `small red edge'-spectra shows one positive band of $V/I = 1.0 \times^{-3}$, where the `large red edge'-spectra shows a similar postive band and a large negative band of $V/I = -4.2 \times^{-3}$. By looking at the trajectory, we identify the small and the large red edge as grass and trees respectively. The circular polarization spectrum averaged over the entire scene would show no clear signature due to the relative large soil (`no red edge') coverage across the scene. 

%-------------------------------------------------------
%-------------------------------------------------------   
%-------------------------------------------------------   
%-------------------------------------------------------
%-------------------------------------------------------
  \begin{figure} [ht!]
  \begin{center}
  \begin{tabular}{c} 
  \includegraphics[width=0.90\textwidth]{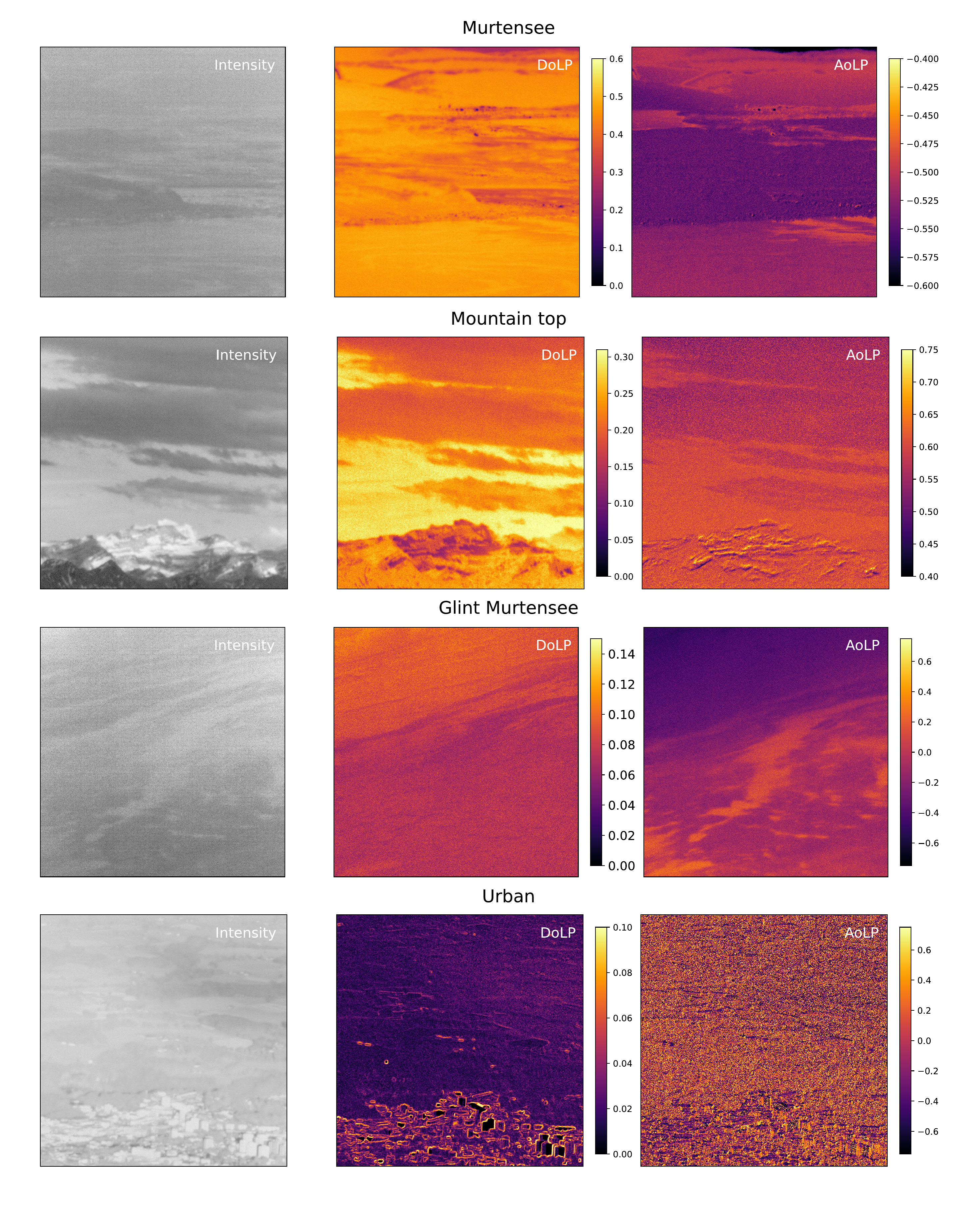}
	\end{tabular}
	\end{center}
  \caption{ [from top to bottom] $600 \times 600$ pixel cut-out of four polarization camera images of (i) the Murtensee and its coast, (ii) a distant mountain top, (iii) the Murtensee featuring a glint and (iv) an urban landscape. [from left to right] Intensity image, DoLP and AoLP, see Equation \ref{eqn: dolp&aolp}, per image.}
  { \label{fig: DoLPAoLP} 
}
  \end{figure} 
%-------------------------------------------------------

From top to bottom, Figure \ref{fig: DoLPAoLP} shows four polarization camera images of the Murtensee and its coast, a distant mountain top, a central point of the Murtensee and an urban landscape. The DoLP reveals structural features of the landscapes, which allows us to distinguish between a mountain top covered in ice and foggy clouds located just above the horizon. For the urban landscape (the middle bottom plot), we clearly distinguish the (white) buildings located in the lower half of the image from the grass fields located in the top half of the image. The AoLP images of the `Murtzensee' and `Glint Murtzensee' reveal a patchy structure on the water surface that we identify as a glint phenomenon.

%%%%%%%%%%%%%%%%%%%%%%%%%%%%%%%%%%%%%%%%%%%%%%%%%%%%%%%%%%%%%%%%%%%%%%%%%%
\section{Conclusions and Discussion}
\label{sec: conclusion}

We described an airborne optical-setup that we used to simultaneously measure circular polarization spectra and broadband DoLP and AoLP from various scenes that include biotic and abiotic features. During the entire flight, reference photos were taken with a regular imaging camera. These encapsulated the area that we were pointing towards, allowing for concise identification of the sources inducing the observed the circular polarization spectra. 
To our knowledge, this was the first time that a hot air balloon was used as an observing platform for spectropolarimetric measurements of the
Earth's surface. We established the maximum solar angle and integration time for which we can obtain circular polarization spectra. This information is of great value for the next upcoming flights. 

At elevations of $\sim$~20~m and $\sim$~650~m (Figures \ref{fig: observation_194857_G0013896} and \ref{fig: observation_200113_G0013921}), we could distinguish between circular polarization spectra of grass and soil. The spectral characteristics of the grass landscapes in this paper are qualitatively similar to those presented by Patty et al. (2021)\cite{Patty21}. We measured a circular polarization of grass of $V/I$ = $2\times10^{-3}$ ($\sim$~20~m elevation) and $V/I$ = $-5\times10^{-4}$ ($\sim$~650~m elevation). Unlike Patty et al. (2021)\cite{Patty21}, we did not observe circular polarization signals from surface water (Figure \ref{fig: observation_201251_G0013943}). A possible explanation is that there is a smaller biomass of photosynthetic organisms in the water, as this study was conducted in the early spring, while Patty et al. 
measured during late summer. 
In addition, the lake was observed in the second half of the balloon flight, close to sunset. The lack of signal could thus also be due to the combination of a large viewing and solar angle. 

We captured how $V/I$ varies for a subsequent observation of farmland featuring various types of grass and possibly dry, and wet soil types 
(Figure \ref{fig: observation_195217_healthyness_vegetation}). The circular polarization spectra for 15 subsequent three-second duration measurements of the farm land reveal the circular polarization ranging between $V/I$ = $-9.0 \times10^{-2} - 0.25 \times 10^{-2}$. We have two explanations for this variation: (i) the arrangement of e.g. crops in soil causes the variation as we interchange observing crops and soil, or (ii) the variation is a consequence of a diversity in health of the observed vegetation. The latter explanation is the more likely one, as we do not see clear crop(-like) fields on the reference photo. Although quantitative circular polarization measurements on the health of the vegetation are lacking, we do know that leaves so show a reduced circular polarization when decaying \cite{Patty17}. Further quantitative and qualitative analysis is required to verify our hypothesis. %vanaf de grond/dichter bij de grond leken heel veel van die velden vrij patchy, ligt natuurlijk aan het gewas maar sowieso zijn die velnden vrij variabel in dichtheid toch

The circular polarization spectra from grass feature solely a single band, often being negative. Unlike the grass spectra, those from tree canopies show vary both from shape and sign. The forest spectra, as presented in Patty et al. (2021)\cite{Patty21}, have both a positive and negative band that exhibit a larger variation in shape and magnitude than the grass spectra. The difference between the two is illustrated in Figure \ref{fig: observation_grass_trees_underdiscussion}. The spectra showing the large negative peak is identified as grass, whereas the single positive peak around 710 nm appears to be induced by tree canopies. During this field campaign, we noticed a high diversity in circular polarization spectra for forests. It remains unclear what mechanisms could cause these variations. Measurements performed on individual (tree) leaves is a first step towards understanding the signals resulting from a canopy. As far as we know, Patty et al. (2022)\cite{Patty22} is the only study that investigated the full-Stokes spectropolarimetry of single leaves from different species and its dependency on the incidence angle of the light source. In their experiment, they varied between a phase angle of 10° to 75° where the phase angle is similar to the angle of the incident light. In general, all their fractional linear ($Q/I$) polarization spectra show a peak around 680 nm due to absorption by chlorophyll and all their fractional circular ($V/I$) spectra feature a negative band around 670 nm and a positive band around 700 nm. The $Q/I$ spectra reveal a strong dependence on phase angle, whereas the $V/I$ spectra are relatively insensitive to changes in the phase angle. 
We are currently investigating the effect of changing the phase angle and viewing angle of multiple-, leaf-to-leaf reflections, see Mulder et al. (in prep). With this knowledge, we will be a step closer to formulate a realistic circular polarization surface model that can be used to accurately simulate various tree canopies and therefore also aid in the interpretation of airborne full-Stokes spectropolarimetric measurements. In addition, these realistic circular polarization surface models would be valuable for realistic Earth-like (exo)planet models\cite{stam08,Groot20}.  

We used the obtained broadband DoLP and AoLP information for landscape identification purposes. It would be interesting to capture linear polarization spectra together with our circular polarization spectra of the surface scenes. This can be achieved by either adding a second spectropolarimeter or a spectropolarimeter that is capable of acquiring full-Stokes polarization from a single data frame \cite{Snik19,Sparks19,Keller20,Mulder21}.  
Peltoniemi et al. (2015)\cite{Peltoniemi15} points out that linear polarization would predominantly provide scalar reflectance information. However, we might be able to obtain more information than surface reflectance, especially when covering multiple phase angles.

Due to the relatively fast ascend and descend, the effect of the balloon elevation on the circular polarized spectra from abiotic and biotic surfaces remains inconclusive. 
Further research (including numerical simulations) will help us to understand the influence of illumination, viewing angles on the circular polarization spectra for various landscapes including high elevation biomes (e.g. tundra, moss, lichens), barren land, possible water bodies (in shallow water) ice, snow, and red algae on snow from different elevations.
Hot air balloons are only able to fly within the timespan of 3 hours before sunset until 3 hours after sunrise. We will plan our next balloon flight in the early morning. This allows us to prepare our instrument during take-off, e.g. taking flat fields, while setting up the polarization and reference camera. Doing so, we will start the scientific measurements just after sunrise, thus preventing large integration times and large scattering angles. In upcoming research, we will include measurements of grass and various tree canopies from one single observation point over the course of an entire day. With this quantitative and qualitative data, we will formulate a realistic circular polarization vegetation model.

%%%%%%%%%%%%%%%%%%%%%%%%%%%%%%%%%%%%%%%%%%%%%%%%%%%%%%%%%%%%%%%%%%%%%%%%%%
\acknowledgments % equivalent to \section*{ACKNOWLEDGMENTS}       

We thank the team of Balloons du Leman (the entire ground team, Julie, pilot Laura and Gael) personally for their enthusiasm, flexibility and help in the preparations for and during the balloon flight. We thank Remko Stuik for the airborne platform brainstorm session.
This work has been carried out within the framework of the NCCR PlanetS supported by the Swiss National Science Foundation under grants 51NF40\_182901 and 51NF40\_205606. This work was supported by the second Planetary and Exoplanetary Science Programme (PEPSci-II) of the Netherlands Organisation for Scientific Research (NWO). 

%%%%%%%%%%%%%%%%%%%%%%%%%%%%%%%%%%%%%%%%%%%%%%%%%%%%%%%%%%%%%%%%%%%%%%%%%%
% References
\bibliography{main} % bibliography data in main.bib
\bibliographystyle{spiebib} % makes bibtex use spiebib.bst

%%%%%%%%%%%%%%%%%%%%%%%%%%%%%%%%%%%%%%%%%%%%%%%%%%%%%%%%%%%%%%%%%%%%%%%%%%
\end{document}